\documentclass[aps,prd,reprint,superscriptaddress,notitlepage,nofootinbib]{revtex4-1}

\usepackage[utf8]{inputenc}
\usepackage[english]{babel}
\usepackage{mathtools}
\usepackage{subdepth}
\usepackage{natbib}
\usepackage[bottom]{footmisc}
\usepackage{kotex}
\usepackage{graphicx}
\usepackage{float}
\usepackage{amssymb}
\usepackage{amsmath}
\usepackage{xcolor}
\usepackage{hyperref}
\usepackage{slashed}
\usepackage{kotex}
\usepackage[shortlabels]{enumitem}
\usepackage{microtype}
\usepackage{comment}
\usepackage{ulem}

\newcommand{\beq}{\begin{equation}}
\newcommand{\eeq}{\end{equation}}

\def\bal{\begin{align}}
\def\eal{\end{align}}

\def\l{\left}
\def\r{\right}

\def\ep{\epsilon}
\def\la{\langle}
\def\ra{\rangle}

\newcommand{\eq}[1]{Eq.~(\ref{#1})}

\newcommand{\GeV}{\mathinner{\mathrm{GeV}}}

\newcommand{\equalcontribmark}{\textsuperscript{}}

\makeatletter
\g@addto@macro\frontmatter@footnote@produce{%
  \begingroup
    \advance\csname c@\@mpfn\endcsname\@ne
    \def\@thefnmark{$\vphantom{x^2}$}%
    \frontmatter@footnotetext{%
      \kern-1em All authors equally contributed to this work.%
    }%
  \endgroup
}
\makeatother

\begin{document}

\title{Primordial black holes as a natural consequence of scale-invariance}

\author{TaeHun Kim\equalcontribmark}
\email[]{gimthcha@kias.re.kr}
\affiliation{School of Physics, Korea Institute for Advanced Study, Seoul 02455, Republic of Korea}
\affiliation{International Center for Quantum-field Measurement Systems for Studies of the Universe and Particles, High Energy Accelerator Research Organization, Ibaraki 305-0801, Japan}
\affiliation{The Institute for Gravitation and the Cosmos, The Pennsylvania State University, University Park, Pennsylvania 16802, United States of America}

\author{Wan-Il Park\equalcontribmark}
\email[]{wipark@jbnu.ac.kr}
\affiliation{Division of Science Education and Institute of Fusion Science, Jeonbuk National University, Jeonju 54896, Republic of Korea}


\begin{abstract}

We show that a scale-invariant $U(1)$ gauge theory can naturally realize large curvature perturbations around the end of thermal inflation, seeding the formation of primordial black holes (PBHs).
The allowed nonminimal gravitational interaction $\xi$ of the Higgs field induces a Hubble-scale negative mass-squared term for positive $\xi \sim \mathcal{O}(1)$, and triggers a second order phase transition that ends thermal inflation while keeping the extra $e$-folds small.
This sharp contrast to the minimally coupled case allows the gauge coupling to be much smaller than unity. 
It turns out that field fluctuations are dominated by thermal fluctuations around the onset of the phase transition, and sufficiently amplified to produce an appreciable amount of PBHs with a wide range of possible masses. 
The associated scalar-induced gravitational waves (GWs) are accompanied by cosmic-string induced GWs coming from the breaking of the $U(1)$ symmetry.
As a result, the generated PBH population can have a correlated GW signature relevant to pulsar timing arrays, space-based interferometers, or ground-based detectors.

\end{abstract}

\pacs{11.10.Kk}
\maketitle

\allowdisplaybreaks


\section{Introduction}

The electroweak hierarchy problem motivates more fundamental scale-invariant extensions of the Standard Model. For anomalous scale symmetries, radiative corrections can dynamically generate a mass scale and lead to the Coleman-Weinberg (CW) type potential.
The plateau nature of the CW potential made it a well-motivated inflationary model in the early stages of development~\cite{Linde:1981mu}, although the minimal realization for primordial inflation was later disfavored by Planck observations~\cite{Planck:2018jri}.

Scale symmetry can remain exact at the quantum level~\cite{Garcia-Bellido:2011kqb,Bezrukov:2012hx,Salvio:2014soa,Ferreira:2016vsc,Ferreira:2016wem}, with the Planck scale arising from its spontaneous breaking. This requires at least one scalar field to quadratically couple to the Ricci scalar, so that its vacuum expectation value (VEV) generates the effective Planck mass. Similar nonminimal interactions are allowed and likely to exist for other scalars, rendering such a nonminimal quadratic gravitational interaction as a general feature of theories with exact scale symmetry. A CW potential can still arise radiatively when the renormalization scale is supplied by a scalar VEV rather than introduced as an explicit mass scale.

Meanwhile, it was shown that thermal inflation~\cite{Lyth:1995hj,Lyth:1995ka} can produce an appreciable abundance of primordial black holes (PBHs) when the flaton potential develops a Hubble scale negative curvature~\cite{Dimopoulos:2019wew}, through spinodal amplification of the field fluctuations and subsequent large density perturbations. Supergravity theories were considered as natural realizations of the negative Hubble-scale mass term, but such theories are likely to require a Planckian VEV of the flaton field, prolonging the thermal inflation. 
This significantly reduces the number of e-folds available for primordial inflation, demanding nontrivial modifications of inflation models from their original forms to match observations.

The CW potential can naturally drive supercooling after primordial inflation, providing a nonsupersymmetric realization of thermal inflation. A related question is then whether the same mechanism can generate perturbations large enough to form PBHs. A study based on a quasi-conformal model has analyzed this possibility~\cite{Lewicki:2021xku}, and found that the curvature perturbations were too small for appreciable PBH production, but the associated scalar-induced gravitational waves (GWs) could reach the NANOGrav sensitivity range~\cite{NANOGRAV:2018hou}.

In this work, we investigate how a nonminimal gravitational coupling $\xi$ of the $U(1)$ Higgs field changes CW-driven thermal inflation. Such a coupling is allowed in ultraviolet (UV) theories with exact scale symmetry and, for positive $\xi$ of order unity, induces a negative Hubble-scale mass term. We show that this term can enhance the curvature perturbations enough to seed PBH formation.~\footnote{
Ref.~\cite{Zhang:2025kbu} also studied PBH formation with a CW-type potential and an additional mass term. The origin of that term and the mechanism ending supercooling differ from those considered here, leading to different parameter spaces and PBH mass ranges.
}

This paper is organized as follows.
Sec.~\ref{sec:model} introduces the scale-invariant local $U(1)$ model and its low-energy potential, and describes thermal inflation which could occur after the primordial inflation.
Sec.~\ref{sec:perturbation} calculates the curvature perturbations generated during the second-order phase transition at the end of the thermal inflation. Sec.~\ref{sec:pheno} discusses PBHs and GWs, and Sec.~\ref{sec:summary} presents our conclusions.

Throughout the paper, we use natural units $c = \hbar = 1$ and the reduced Planck mass $M_{\rm P} = 1/\sqrt{8\pi G}$.

\section{A model with scale-invariance}
\label{sec:model}

\subsection{A scale-invariant $U(1)$-gauge theory}

We consider a scale-invariant $U(1)$ gauge theory with a complex scalar $\tilde\Phi$, where a tilde denotes a Jordan-frame quantity. After spontaneous breaking of scale symmetry generates the Planck scale~\cite{Ferreira:2016vsc,Ferreira:2016wem}, the low-energy action takes the form
\beq
S = \int d^4 x \sqrt{-{\tilde g}} \l( \mathcal{L}_{\rm g} + \mathcal{L}_{\rm m} \r)
\eeq
where ${\tilde g}$ is the determinant of the Jordan frame metric tensor ${\tilde g}_{\mu \nu}$, and
\begin{align}
\mathcal{L}_{\rm g} &= - \frac{1}{2} \Omega^2({\tilde \Phi}) M_{\rm P}^2 {\tilde R}
\\
\mathcal{L}_{\rm m} &= {\tilde g}^{\mu \nu} \l( D_\mu {\tilde \Phi}^* \r) \l( D_\nu {\tilde \Phi} \r) - {\tilde V}({\tilde \Phi})
\end{align}
with $D_\mu = \partial_\mu + ig A_\mu$ being the covariant derivative.
For $|{\tilde \Phi}| \ll M_{\rm P}$, the frame function can be approximated as
\beq
\Omega^2({\tilde \Phi}) \equiv 1 + \frac{\xi |{\tilde \Phi}|^2}{M_{\rm P}^2} \, , 
\eeq
where dimensionless coupling $\xi$ can have either sign but we take $\xi > 0$ in this work. The potential is
\beq
{\tilde V}({\tilde \Phi}) = V_0 + \lambda |{\tilde \Phi}|^4 \, ,
\eeq
where the coupling $\lambda$ is taken to be positive.
A nonzero $V_0$ is compatible with scale-invariant theories once spontaneous symmetry breaking generates mass scales.

Writing the Jordan-frame radial field as ${\tilde \phi} \equiv \sqrt{2} |{\tilde \Phi}|$, the Weyl rescaling
\beq
g_{\mu \nu} \equiv \Omega^2({\tilde \phi}) {\tilde g}_{\mu \nu}
\eeq
leads to the Einstein-frame action
\beq
S = \int d^4 x \sqrt{-g} \l[ - \frac{1}{2} M_{\rm P}^2 R + \frac{1}{2} \l(\partial \phi \r)^2 - V^{(0)}(\phi) \r] \, ,
\eeq
where 
\beq
\phi = \int_0^{\tilde \phi} F({\tilde \phi}') d {\tilde \phi}'
\eeq
with 
\beq
F({\tilde \phi}) = \Omega({\tilde \phi})^{-1} \sqrt{1+6 M_{\rm P}^2 \l( d \Omega({\tilde \phi})/ d {\tilde \phi} \r)^2} \, , 
\eeq
and $V^{(0)}(\phi) \equiv \tilde V ({\tilde \phi})/\Omega^4({\tilde \phi})$.
For ${\tilde \phi} \ll M_{\rm P}$ and $\xi \sim \mathcal{O}(1)$, $F \simeq 1$ and $\tilde \phi \simeq \phi$ \footnote{The next leading term, $\xi\l( 1 - 3\xi \r) \phi^3/12 M_{\rm P}^2$, does not affect our argument, and hence was neglected.}.
Then, the tree-level Einstein frame potential can be approximated as
\beq
V^{(0)}(\phi) \simeq V_0 - \frac{\xi V_0}{M_{\rm P}^2} \phi^2 + \frac{\lambda}{4} \phi^4 \, .
\eeq
Thus, during the vacuum domination by $V_0$, $\phi$ receives a constant Hubble-scale negative mass-squared for $\xi \sim \mathcal{O}(1)$.

We assume the symmetry-breaking radiative corrections are dominated by gauge interactions. This results in the one-loop zero-temperature CW potential
\beq \label{eq:V-1loop}
V(\phi) = V_0 - \frac{\xi V_0}{M_{\rm P}^2} \phi^2 + \frac{3g^4}{64 \pi^2} \phi^4 \l[ \ln \l( \frac{\phi^2}{v_\phi^2} \r) - \frac{1}{2} \r] \, , 
\eeq
where we neglect loop corrections to $\xi$, and $v_\phi$ is the symmetry-breaking VEV. To be consistent with the negligible vacuum energy of the present universe, we set
\beq
V_0 \simeq \frac{3 g^4 v_\phi^4}{128 \pi^2} \, .
\eeq

\subsection{Thermal inflation and phase transition}
In the early universe with finite temperature, the temperature-dependent contributions are added to 
\eq{eq:V-1loop}.
In the high temperature limit, the effective potential in the region of $\phi \lesssim T/g$ can be approximated as
\begin{align} \label{eq:VT-eff}
V(\phi, T) \approx{}& V_0 - \frac{\pi^2}{10} T^4 - \frac{\xi V_0}{M_{\rm P}^2} \phi^2 + \frac{g^2}{8}T^2 \phi^2
\nonumber \\
&- \frac{g^3}{4\pi}T\phi^3 + \frac{3 g^4}{64 \pi^2} \phi^4 \l[ \ln \l( \frac{\phi^2}{v_\phi^2} \r) - \frac{1}{2} \r] \, .
\end{align}
For simplicity, we neglect Daisy contributions \cite{Dolan:1973qd}, which give only minor changes to the details of the phase transition discussed subsequently.

If the reheating temperature after the primordial inflation was sufficiently high, the thermal corrections could trap the Higgs field $\phi$ at the origin.
In such a case, thermal inflation begins when the vacuum energy at the origin, $V_0$, starts dominating over the radiation energy density.
At this epoch, the temperature is 
\beq \label{eq:Tb}
T_{\rm b} \approx \l( \frac{\pi^2 g_*(T_{\rm b})}{30} \r)^{-1/4} V_0^{1/4} \, ,
\eeq
where $g_*(T)$ is the number of relativistic degrees of freedom at temperature $T$ of the thermal background, and the expansion rate is $H_{\rm b} \approx \sqrt{2 V_0}/\sqrt{3} M_{\rm P}$. The time and the scale factor at this moment are denoted by $t_b$ and $a_b$, respectively.

\begin{figure}[ht]
    \centering
    \includegraphics[width=0.98\linewidth]{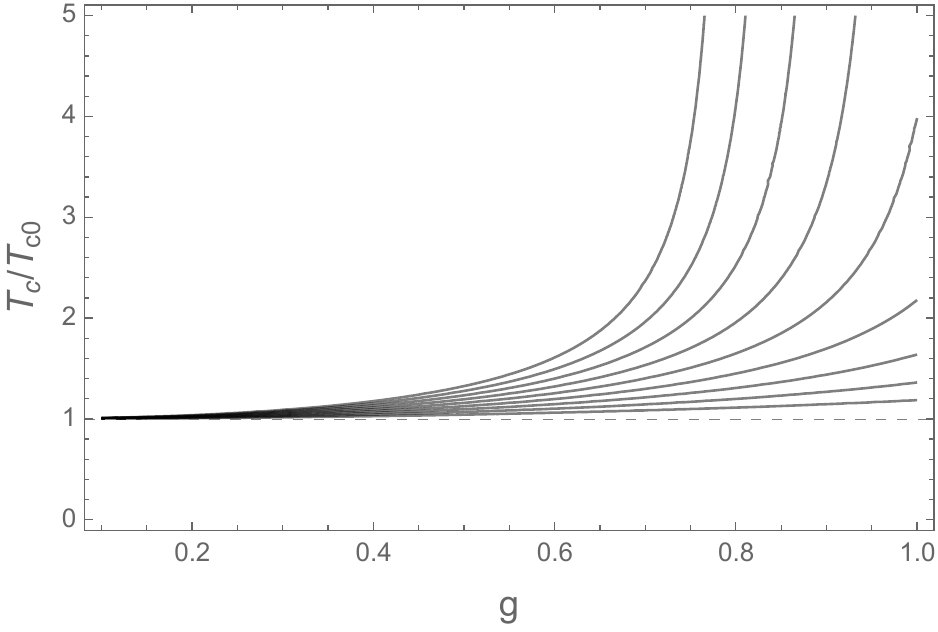}
    \caption{The dependence of $T_{\rm c}/T_{\rm c 0}$ on $g$ for $\xi=1.0$ and $v_\phi=10^i {\rm GeV}$ with $i = (0, 2, 4, ..., 16)$ from left to right lines.  
    As long as $\xi = \mathcal{O}(1)$, the impact of $\xi$ is minor as can be seen from \eq{eq:Tc}. 
    }
    \label{fig:Tc-to-Tc0}
\end{figure}

As the Universe cools down, thermal inflation ends by either a first or second order phase transition. The phase transition happens around or at the moment when the effective mass squared at the origin $m_{\rm eff}^2$ vanishes. We denote the temperature, time, and scale factor at this moment by $T_c$, $t_c$, and $a_c$, respectively.
From \eq{eq:VT-eff}, keeping the leading logarithmic contribution in the cactus approximation gives the estimate
\begin{align} \label{eq:Msq}
m_{\rm eff}^2 &\equiv \la V''(\delta \phi, T) \ra_{\phi=0} 
\nonumber \\
&\approx \frac{g^2}{4} \l( T^2 - T_{\rm c0}^2 \r) - \frac{9 g^4}{16 \pi^2} \la \l( \delta \phi \r)^2 \ra \ln \l( \frac{v_\phi^2}{\la \l( \delta \phi \r)^2 \ra} \r) \, , 
\end{align}
where 
\beq \label{eq:Tc0}
T_{\rm c0} \equiv \frac{\sqrt{24 \xi} H_{\rm c}}{g} = \frac{\sqrt{3 \xi} g v_\phi^2}{4 \pi M_{\rm P}}
\eeq
with $H_{\rm c} \simeq V_0^{1/2} / \sqrt{3} M_{\rm P}$ being the expansion rate around the end of thermal inflation.
The vacuum and thermal fluctuations of $\phi$ at $T\sim T_{\rm c}$ are given by \cite{Vilenkin:1982nj}
\begin{align}
\label{eq:vac-fluc}
\la (\delta \phi)^2 \ra_{\rm vac} &\approx \l( \frac{H_{\rm c}}{2\pi} \r) ^2 \, ,
\\ 
\label{eq:thermal-fluc}
\la (\delta \phi)^2 \ra_T &\approx \frac{T^2}{12} + \frac{H_{\rm c}^2}{4 \pi g} \, .
\end{align} 
The thermal fluctuation is dominant for the parameter space on which we focus, $g \lesssim \mathcal{O}(0.1)$ and $\xi \sim \mathcal{O}(1)$.
We approximate $\la \l(\delta \phi \r)^2 \ra \approx T_{\rm c0}^2/12$ in the log factor in \eq{eq:Msq} as $T$ approaches $T_{\rm c}$.
Then, with $\bar{\lambda}_{\rm c} \equiv ( 3g^4 / 16 \pi^2 ) \ln ( v_\phi^2/\la \l( \delta \phi \r)^2 \ra )$, the critical temperature for $m_{\rm eff}^2=0$ is found to be
\beq \label{eq:Tc}
T_{\rm c} = \l[ \frac{1 + \bar{\lambda}_{\rm c} / 8 \pi g \xi}{1 - \bar{\lambda}_{\rm c}/g^2} \r]^{1/2} T_{\rm c0} \approx T_{\rm c0} \, ,
\eeq
where the approximation is valid for $\bar{\lambda}_{\rm c} \ll g^2$ which is well satisfied for $g \lesssim \mathcal{O}(0.1)$. 
Note that Eq.~\eqref{eq:Tc} ceases to be reliable when $\bar{\lambda}_c / g^2$ approaches unity. For a given $v_\phi$, $\bar{\lambda}_c$ becomes effectively a function of $T$ as long as $T \gtrsim H_{\rm c}/\sqrt{g}$, and such a breakdown can happen during $T > T_c$ if $g$ is large enough. The exact dependence of $T_c$ on $g$ and $v_\phi$ is plotted in Fig.~\ref{fig:Tc-to-Tc0}, and we focus only on cases with $\bar{\lambda}_{\rm c}/g^2 <1$ with $\xi>0$.~\footnote{The case with $\xi<0$ is expected to share some properties of the phase transition with the case discussed in Refs.~\cite{Lewicki:2021xku,Zhang:2025kbu}, although we do not discuss this case here.
}

\begin{figure}[ht]
    \centering
    \includegraphics[width=0.98\linewidth]{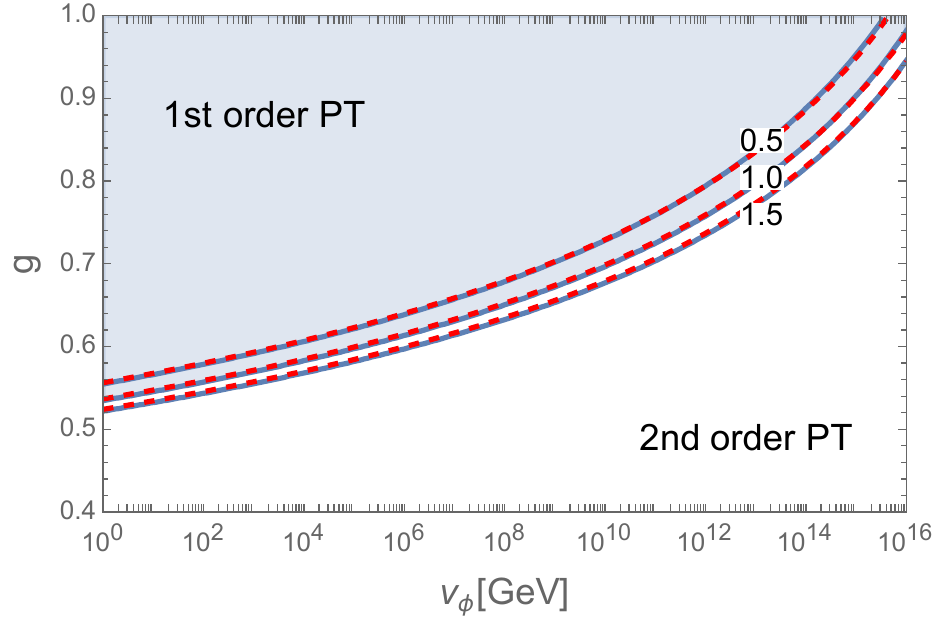}
    \caption{
    The parameter space for each type of phase transition in the $(v_\phi,g)$ plane for $\xi=0.5,1,$ and $1.5$, from top to bottom. Solid curves use the numerical bounce action obtained with \textbf{CosmoTransitions}, and dashed curves use the analytic approximation in Appendix~\ref{app:PT}, showing excellent agreement. The regions above and below each curve are classified as first-order and second-order transitions, respectively, under the criterion specified in Appendix~\ref{app:PT}.
    }
    \label{fig:PT-para-space}
\end{figure}

The nature of the transition depends on $\xi$, $g$, and $v_\phi$, as shown in Fig.~\ref{fig:PT-para-space}.
We also found an analytic approximation of the bounce action and present it in App.~\ref{app:PT}.
Relative to the minimally coupled case~\cite{Lewicki:2021xku}, the negative Hubble-scale mass term shifts the critical value of $g$ separating different types of phase transitions toward larger values, while having qualitatively similar dependence on $v_\phi$.
This is because satisfying the bubble-nucleation condition requires a larger thermal correction when the nonminimal coupling adds a negative contribution to the effective mass squared.
In the present work, we focus on the second order phase transition as it can produce larger density perturbations than the first order case. We thus extend the range of $g$ to values well below the critical value in our numerical study. The numerical results in the first-order region of Fig.~\ref{fig:PT-para-space} should be understood as a formal continuation of the assumed second-order case.

The total duration of thermal inflation $\Delta N_{\rm TI}$ is a sum of contributions from the near de Sitter phase and the subsequent fast-roll phase:
\beq
\Delta N_{\rm TI} = \Delta N_{\rm ds} + \Delta N_{\rm fr} \, .
\eeq
The near de Sitter phase spans
\begin{align}
\Delta N_{\rm ds} 
&\approx \ln \frac{T_{\rm b}}{T_{\rm c}} 
\nonumber \\
&= \frac{1}{4} \ln \l[ \frac{20}{g_*(T_{\rm b}) \xi^2} \l( \frac{1-\bar{\lambda}_{\rm c}/g^2}{1+\bar{\lambda}_{\rm c}/8 \pi g \xi} \r)^2\r] + \ln \l( \frac{M_{\rm P}}{v_\phi} \r)
\nonumber \\
&\simeq 42.0 -\frac{1}{2} \ln \xi - \ln \l( \frac{v_\phi}{1 {\rm GeV}} \r) \, , 
\end{align}
where $\bar{\lambda}_{\rm c}/g^2 \ll 1$ and $g_*(T_{\rm b})=100$ were taken in the last line. 

The fast-roll phase mainly happens in $\phi \lesssim \phi_\times$, where the negative mass term dominates the dynamics of $\phi$.
Equating the magnitudes of the derivatives of the quadratic and CW terms gives
\beq \label{eq:phi-x}
\phi_\times 
= \frac{4\pi \sqrt{\xi}}{g^2 \ln^{1/2} \l( v_\phi/\phi_\times \r)} H_{\rm c} \, .
\eeq
The impact of the contribution from the cactus approximation on $\phi_\times$ turned out to be minor and does not affect our result for the whole range of parameter space of $g < 1$ and $v_\phi \gtrsim 1 \GeV$, hence was neglected.

\begin{figure}[ht]
    \centering
    \includegraphics[width=0.98\linewidth]{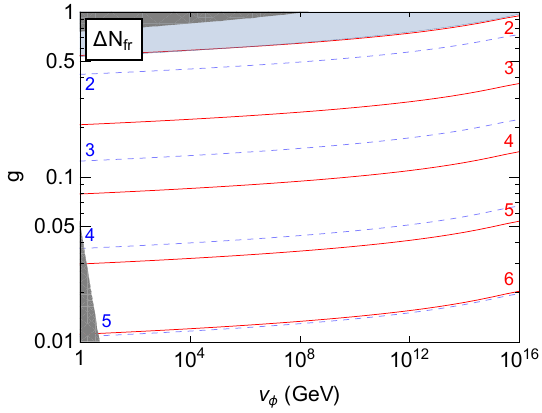}
    \includegraphics[width=0.98\linewidth]{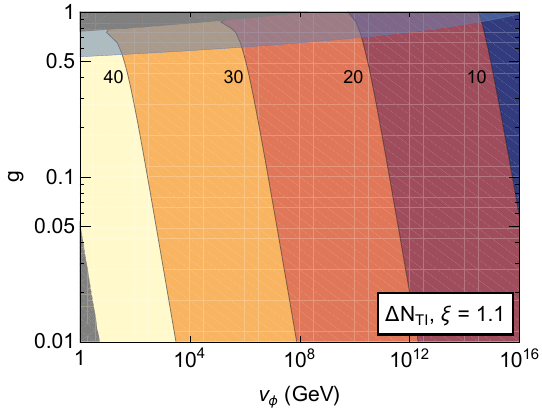}
    \caption{\textit{Upper:} Numerically obtained fast-roll $e$-folds for $\xi=1.1$ (red) and $1.5$ (blue dashed).
    \textit{Lower:} The total $e$-folds of thermal inflation as a function of $g$ and $v_\phi$ for $\xi=1.1$. 
    The contours correspond to $\Delta N_{\rm TI} = 10, 20, 30, 40$ from right to left. 
    In both panels, the lower-left gray shaded region has a reheating temperature of thermal inflation below $5 \, \text{MeV}$ and is excluded by the BBN requirement. The upper-left gray shaded region lies outside the validity of the cactus approximation in Eq.~\eqref{eq:Tc}. The blue shaded region is the parameter space of first order phase transition for $\xi = 1.1$.
    }
\label{fig:dN-frNum}
\end{figure}

The end of thermal inflation may be analytically approximated by the epoch of $\la \phi^2 \ra^{1/2} = \phi_\times$.
The dynamical time scale of the field modes during the tachyonic stage with $\phi \lesssim \phi_\times$ is $\tau_{\rm dyn} \sim 1/H_{\rm c}$, whereas the time scale of thermalization is $\tau_{\rm th} \gtrsim 1/ g^4 T_{\rm c} \approx 1/\sqrt{24\xi} g^3 H_{\rm c}$.
So for $g \lesssim \mathcal{O}(0.1)$, we expect $\tau_{\rm dyn} \ll \tau_{\rm th}$, and the stochastic effect of thermal bath on the field dynamics could be ignored once the tachyonic instability is triggered~\cite{Rosa:2021gbe}.
Then, the evolution of superhorizon modes $\phi_{k, L}$ becomes
\beq
\phi_{k, L}(t) \approx \phi_{k, L}^i \exp \l[ \bar{\nu} H_{\rm c} \l(t-t_{\rm c} \r)\r] \, , \label{eq:phikL}
\eeq
where $\phi_{k, L}^i$ is the initial value and
\beq
\bar{\nu} \equiv \frac{3}{2} \l[ \sqrt{1+\frac{4m_0^2}{9H_{\rm c}^2}} - 1 \r] = \frac{3}{2} \l[ \sqrt{1+ \frac{8\xi}{3}} - 1 \r]
\eeq
with $m_0 \equiv \sqrt{6\xi} H_{\rm c}$.
Further applying the approach in Ref.~\cite{Fonseca:2010nk} with the thermal fluctuations of \eq{eq:thermal-fluc} as subhorizon perturbations at the end of de Sitter phase, we find the extra $e$-folds from the fast-roll phase as
\begin{align} \label{eq:dN-fr}
\Delta N_{\rm fr}
&\approx \frac{1}{2 \bar{\nu}} \l[ \ln \l( \frac{\phi_\times^2}{\la \l( \delta \phi \r)^2 \ra_T} \r) + \ln 2 + \gamma_{\rm E} - \frac{\la \l( \delta \phi \r)^2 \ra_{\rm vac}}{\la \l( \delta \phi \r)^2 \ra_T} \r]
\nonumber \\
&\approx \frac{1}{\bar{\nu}} \l[ \ln \l( \frac{2 \sqrt{2} \pi}{g \ln^{1/2} \l( M_{\rm P}/v_\phi \r)} \r) + \frac{1}{2} \l( \ln 2 + \gamma_{\rm E} \r) \r] \, , 
\end{align}
where only the leading contribution was taken in the second line with an approximation $\phi_\times \approx v_\phi^2/M_{\rm P}$ for the log-factor in \eq{eq:phi-x}.
Fig.~\ref{fig:dN-frNum} shows the numerical estimate of $\Delta N_{\rm fr}$, and is qualitatively consistent with \eq{eq:dN-fr}. 
The number of $e$-folds from the rolling-out in the region $\phi \gtrsim \phi_\times$ turns out to be $\sim \mathcal{O}(0.1)$, although we include this in all our numerical results, including $\Delta N_{\rm fr}$ here and PBH mass and GW frequency in Sec.~\ref{sec:pheno}. 

For $\bar\lambda_c/g^2\ll1$, $\Delta N_{\rm ds}$ depends only weakly on $g$. $\Delta N_{\rm fr}$ also only logarithmically depends on $g$ because the tachyonic mass is Hubble scale. Consequently, in our scenario, small $g$ values do not necessarily lead to an excessively long thermal inflation, leaving room for primordial inflation to generate the observed large-scale perturbations.

\section{Cosmological perturbations across the phase transition} \label{sec:perturbation}

Once the symmetric configuration at the origin becomes unstable, the field in each correlated region evolves away from the origin along a locally selected field-space direction. The ensemble mean of the original complex field remains zero, $\langle \tilde \Phi \rangle = 0$, while its variance is nonzero and grows, $\langle \phi^2 \rangle \neq 0$. During this phase transition, spinodal amplification of these fluctuations generates the curvature perturbations, and we calculate them in this section.

In our scenario, when the second-order phase transition starts, thermal fluctuations dominate over the vacuum fluctuations.
We evaluate the curvature perturbations at the end of the quadratic regime, $\phi\sim\phi_\times\ll v_\phi$, since the phase transition quickly ends as soon as $\phi$ goes beyond $\phi_\times$. Near the onset of the transition, the cactus correction is small for $g \lesssim \mathcal O (0.1)$ as in Fig.~\ref{fig:Tc-to-Tc0}, and the effective mass squared is approximately
\beq \label{eq:msq-eff}
m_{\rm eff}^2 \approx - m_0^2 \l[ 1 - \l( \frac{a_{\rm c}}{a} \r)^2 \r] \, .
\eeq
Here we disregard the exceptional region with $\bar{\lambda}_c / g^2 \geq 1$ for consistency. The temperature dependence of $m_{\rm eff}^2$ allows an analytic solution for the vacuum mode functions~\cite{Dimopoulos:2019wew}, which we use together with the thermal occupation number $n_k$ to specify the initial fluctuations~\cite{Lewicki:2021xku}. We then directly evolve the power spectrum in the subsequent tachyonic phase, with the cactus correction restored.

To obtain the vacuum fluctuations, we rescale the field as $\chi \equiv a \, \phi$ and use the Fourier expansion~\footnote{We follow the convention of Ref.~\cite{Dimopoulos:2019wew}}
\begin{equation}
    \hat{\chi}(\vec{x}, t) = \int \frac{d^3 k}{(2\pi)^{3/2}} \left[ \hat{a}_{\vec{k}} \chi_k (t) e^{i \vec{k} \cdot \vec{x}}  + \text{h.c.} \right] \, , \label{eq:chihatFourier}
\end{equation}
where the ladder operators satisfy the commutation relation $[\hat{a}_{\vec{k}} , \hat{a}_{\vec{k}'}^\dagger] = \delta^3 (\vec{k} - \vec{k}')$. 
In conformal time $\eta = -1/a H_c$, the mode function $\chi_k$ follows the equation of motion
\begin{equation}
    \chi''_{k} + \left(k^2 + a_{\rm c}^2 m_0^2 - \frac{m_0^2 / H_c^2 + 2}{\eta^2} \right) \chi_k = 0 \, , \label{eq:chimodfuncEOM}
\end{equation}
where $k$ is the comoving wave number. 
We hereafter normalize the scale factor as $a_{\rm c} = 1$, so that Eq.~\eqref{eq:chimodfuncEOM} becomes the same as the case of a constant mass squared $-m_0^2$ after the substitution $k^2 \rightarrow k^2 + m_0^2$. 
The mode function solution is found as
\begin{equation}
    \chi_k (\eta) = -\frac{1}{2} \sqrt{-\pi \eta} \, H^{(1)}_{\nu} \left(-\sqrt{k^2 + m_0^2} \, \eta \right) \, , \label{eq:chimodfunc}
\end{equation}
where $H^{(1)}_{\nu}$ is the Hankel function of the first kind and $\nu^2 \equiv 9/4 + m_0^2 / H_c^2 = (\bar{\nu}+3/2)^2$. 

In the stable regime, thermal fluctuations are described by the occupation number
\begin{equation}
    n_k = \left[e^{\sqrt{k^2 - m_0^2 (a^2 - 1)} \, / \, T_{\rm c}} - 1 \right]^{-1} \, . \label{eq:nk}
\end{equation}
But during the second order phase transition, $m_{\rm eff}^2 < 0$ and sufficiently low-$k$ modes have imaginary frequencies. Eq.~\eqref{eq:nk} is thus unsuitable for these modes, yielding complex values of $n_k$. 
However, the field fluctuations themselves remain well defined; it is the particle interpretation that fails for imaginary frequencies. We therefore work directly with the field fluctuations through a Langevin-like stochastic differential equation. Eqs.~\eqref{eq:chimodfunc} and \eqref{eq:nk} then determine the initial thermal fluctuations of the tachyonic stage.

\subsection{Evolution of power spectrum of $\phi$} \label{sec:phipowerspectrum}

For a generic quadratic potential, Fourier modes of the field evolve independently. With the Gaussian approximation of the thermal noise, their covariances of the field and velocity satisfy a closed deterministic system. Through the procedure outlined in App.~\ref{app:phipowerspectrum}, we obtain the corresponding third-order deterministic differential equation for the power spectrum as Eq.~\eqref{eq:PkEOMGeneral}.

Meanwhile, the concave-downward potential in the tachyonic regime $T < T_c$ causes rapid growth of $\phi$, hindering its thermal equilibrium with the bath especially for small coupling $g$. On the other hand, the dynamics of $\phi$ is already dominated by the tachyonic growth. We therefore approximate the subsequent evolution by neglecting the bath-induced thermal noise and dissipation (see the discussion above Eq.~\eqref{eq:phikL})~\cite{Rosa:2021gbe}, only retaining the temperature dependence of the potential and initial thermal fluctuations described by Eqs.~\eqref{eq:chimodfunc} and \eqref{eq:nk}. 
Then, Eq.~\eqref{eq:PkEOMGeneral} reduces to
\begin{equation} \label{eq:PkEOM}
    \dddot{P}_k + 9 H_c \ddot{P}_k + (18 H_c^2 + 4 q_k^2) \dot{P}_k + [2 (q_k^2)^{\cdot} + 12 H_c q_k^2] P_k = 0 \, ,
\end{equation}
where $P_k$ is the power spectrum defined by
\begin{equation}
    \langle \phi_{\vec{k}} \phi_{\vec{k}'}\rangle \equiv \delta^3(\vec{k} + \vec{k}') P_k \label{eq:powerspectrumdefine}
\end{equation}
and we approximated $H(t) \simeq H_c$. 

As the fluctuations approach $\phi_\times$, the cactus contribution becomes important even if it was negligible in Eq.~\eqref{eq:msq-eff}. We therefore evolve Eq.~\eqref{eq:PkEOM} with the effective frequency
\begin{equation}
    q_k^2(t) = \frac{k^2}{a^2(t)} + m_{\rm eff}^2 (t) \, , \label{eq:qksq}
\end{equation}
with $m_{\rm eff}^2 \simeq \la V''(\delta \phi, T) \ra_{G, \phi=0}$ from Eq.~\eqref{eq:VT-eff} (with $\phi \rightarrow |\phi|$ for the cubic term) using the leading Gaussian state~\cite{Cormier:1998nt, Kleinert:1998vv, Cormier:1999ia, Richard:2015iza}
\begin{equation}
    \langle \hat{\mathcal{O}}(\hat{\phi}) \rangle_G = \int^\infty_{-\infty} d\phi \, \frac{1}{(2\pi \langle \hat{\phi}^2 \rangle)^{1/2}} \, e^{-\phi^2 / 2 \langle \hat{\phi}^2 \rangle} \, \mathcal{O}(\phi) \, . \label{eq:VppG}
\end{equation}
The variance is self-consistently obtained from the power spectrum by
\begin{equation}
    \langle \hat{\phi}^2 \rangle(t) = \int^{a(t) H_c}_{a_b H_c} dk\,  \frac{k^2}{2\pi^2} P_k (t) \, . \label{eq:phisqexpval}
\end{equation}
In the numerical evolution, we use an analytic expression for the Gaussian integral for $m_{\rm eff}^2$ while numerically updating the momentum integral in Eq.~\eqref{eq:phisqexpval}. We used the $e$-folds as the time variable.

Eq.~\eqref{eq:PkEOM} requires initial conditions $P_k (t_c)$, $\dot P_k (t_c)$, and $\ddot P_k (t_c)$. We adopt a freeze-out prescription in the Kibble-Zurek mechanism, in which the thermal contribution ceases to track equilibrium from the moment $t_* < t_c$, when the relaxation time equals the quenching time in Eq.~\eqref{eq:quenching}. 
For $t < t_*$, the thermal and vacuum spectra are related by $P_k^{\rm thm}=2n_kP_k^{\rm vac}$, with $P_k^{\rm vac}=|\chi_k|^2/a^2$. The thermal fluctuation is frozen in $t_* < t < t_c$, while the vacuum modes continue to evolve according to Eq.~\eqref{eq:chimodfunc}. The initial condition is then 
\begin{align}
    P_k (t_c) &= P_k^{\rm thm} (t_*) + P_k^{\rm vac} (t_c) \, , \\
    \dot{P}_k (t_c) &= \dot{P}_k^{\rm vac} (t_c) \, ,  \\
    \ddot{P}_k (t_c) &= \ddot{P}_k^{\rm vac} (t_c) \, .
\end{align}
For $t > t_c$, both the thermal and vacuum fluctuations evolve by the tachyonic growth.

\subsection{Curvature perturbations} \label{ssec:curvpert}

The power spectrum of field fluctuations can be converted to the power spectrum of curvature perturbation $\zeta$. 
Had $\phi$ been a conventional single-field slow-roll inflaton, the classical zero mode $\langle \phi \rangle$ would give the curvature perturbation in spatially flat gauge by $\zeta \simeq H (\delta \phi / \partial_t \la \phi \ra)$. In our case, $\phi$ is the radial field defined by $\phi = \sqrt{2} |\Phi|$, so $\langle \phi \rangle \neq 0$ always except at the origin, even though $\langle \Phi \rangle = 0$ by the symmetry. We thus can treat $\phi$ as an ordinary scalar field, and use $\langle \phi^2 \rangle^{1/2}$ to define the background dynamics in the usual way for primordial inflation.

Curvature perturbations from thermal inflation can be calculated through \cite{Dimopoulos:2019wew, Lewicki:2021xku}
\begin{equation} \label{eq:zeta}
    \zeta = H \frac{\delta \rho}{\dot{\rho}} \simeq H \frac{m_{\rm eff}^2 \, \delta (\phi^2)}{\partial_t (m_{\rm eff}^2 \langle \hat{\phi}^2 \rangle)}
   \simeq H \frac{\delta (\phi^2)}{\partial_t \langle \hat{\phi}^2 \rangle}
\end{equation}
with $\rho \simeq \la V(\phi) \ra$ taken as the energy density, and the power spectrum of $\phi^2$ is obtained by
\begin{equation}
    P_{\delta(\phi^2), k} = 2 \int \frac{d^3 \vec{p}}{(2\pi)^3} P_p P_{|\vec{k} - \vec{p}|} \, .
\end{equation}
The last approximation in \eq{eq:zeta} holds for $N_{\rm fr} \gtrsim 1$ as $m_{\rm eff}^2 \approx - m_0^2$, and agrees with the $\delta \mathcal{N}$ formalism adapted for hybrid inflation models~\cite{Fonseca:2010nk, Gong:2010zf}. 
The curvature power spectrum can then be calculated as
\begin{equation} \label{eq:zetapowerspec}
    P_{\zeta , k} \simeq \frac{1}{(d \langle \hat{\phi}^2 \rangle / dN)^2} \, P_{\delta(\phi^2), k} \, .
\end{equation}

Within the cactus approximation in $\phi < \phi_\times$, the higher order interaction terms in the potential give corrections to the effective mass, leaving the separability of Fourier modes valid. 
This preserves the Gaussianity of $\phi$ under our treatment, providing the basis for the Gaussian-field calculation hereafter. 
Nevertheless, the conversion in Eq.~\eqref{eq:zeta} induces nonGaussianity for $\zeta$. $\delta\mathcal{N}$ formalism estimates $f_{\rm NL}^{\rm local} \sim \mathcal{O}(1)$ and impacts PBH formation, however, this can be compensated by a moderate change of $\xi$. Therefore, we stick to the Gaussian approximation implemented in previous studies~\cite{Dimopoulos:2019wew, Lewicki:2021xku} to estimate the subsequent PBH formation and GW spectra, and focus on demonstrating the effect of nonminimal coupling by comparing with those works consistently. 
A detailed treatment of the nonGaussian probability tail and corrections to the phenomenology is left for future work.

\begin{table}[t]
    \centering
    \begin{tabular}{ |c| c| c| }
        \hline
        Name & $(g, \, v_\phi/\text{GeV}, \, \xi)$ & $(M_{\rm PBH} / M_\odot, \, f_{\rm PBH})$ \\ \hline \hline
        SMBH & $(0.1, \, 2.2\times10^1, \, 1.1)$ & $(6.3 \times 10^1, \,  3.7\times 10^{-8})$ \\ \hline
        DM & $(0.1, \, 5.0\times 10^8, \, 1.1)$ & $(1.8\times 10^{-13}, \,  8.4\times10^{-1})$  \\ \hline 
        Evap. & $(0.1, \, 1.0\times 10^{15}, \, 1.1)$ & $(8.6\times 10^{-26}, \,  2.6 \times 10^6)$ \\ \hline 
    \end{tabular}
    \caption{Benchmark parameters and peak PBH masses and abundances for PBH populations giving SMBH seeds, dark matter (DM), and evaporating PBHs. Each benchmark corresponds to one of the three X markers in Fig.~\ref{fig:PBH}.}
    \label{tab:bench}
\end{table}

\begin{figure}[t]
    \centering
    \includegraphics[width=0.98\linewidth]{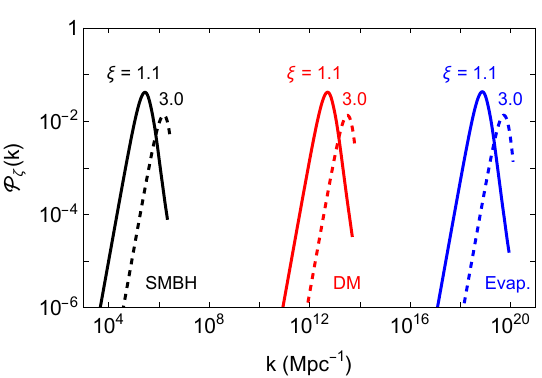}
    \caption{Dimensionless curvature power spectra for $v_\phi = 2.2\times 10^{1} \, \text{GeV}$ (black), $5.0\times 10^{8} \, \text{GeV}$ (red), and $1.0 \times 10^{15} \, \text{GeV}$ (blue), with $g = 0.1$ and $\xi = 1.1$ (solid) and $\xi = 3$ (dashed). The selected sets of parameters correspond to three X markers in Fig.~\ref{fig:PBH} and are summarized in Table~\ref{tab:bench}.
    }
    \label{fig:powerspectrum}
\end{figure}

\begin{figure}[t]
    \centering
    \includegraphics[width=0.98\linewidth]{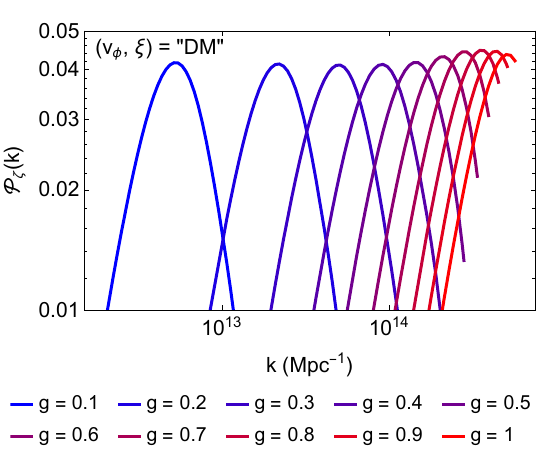}
    \caption{Dimensionless curvature power spectra near the peak for $0.1 \leq g \leq 1$, with $v_\phi= 5.0 \times 10^{8} \, \text{GeV}$ and $\xi = 1.1$ from the ``DM'' benchmark point.}
    \label{fig:powerspectrumvaryingg}
\end{figure}

We evaluate the dimensionless curvature power spectrum
\begin{equation} \label{eq:dimlesszetapowerspec}
    \mathcal{P}_{\zeta, k} \equiv \frac{k^3}{2\pi^2} P_{\zeta, k}
\end{equation}
for three benchmark points related to PBH formation listed in Table~\ref{tab:bench}, with an additional value of $\xi$ for comparison.
The result is shown in Figs.~\ref{fig:powerspectrum} and \ref{fig:powerspectrumvaryingg}.
Since the potential in the quadratic regime is the same as the potential considered in Ref.~\cite{Dimopoulos:2019wew}, its analytic estimate of the curvature power spectrum, Eq.~(2.24), provides a useful guide to the parameter dependence.

The peak lies near modes that exit the horizon shortly after $t_c$. 
If only vacuum fluctuations were used, the characteristic peak scale would be $k_{\zeta, \rm peak} / H_c = \sqrt{3(\bar{\nu}+3)/2} \sim \mathcal{O}(1)$~\cite{Dimopoulos:2019wew}. Although thermal fluctuations dominate the spectrum around $t_c$ in our scenario, numerical results show that the ratio $k_{\zeta, \rm peak} / H_c$ is not much affected, only slightly increased.
PBH formation and scalar-induced GWs inherit this characteristic scale, and their spectra are also peaked.

The curves end at the UV cutoff we imposed, in which the field beyond this point quickly rolls down. We expect the power spectrum to be suppressed, and do not alter the main result of the present work. The infrared cutoff corresponds to modes crossing the horizon at $t_b$, the beginning of thermal inflation, and lies outside the plotted range.

At fixed $g$ and $\xi$, increasing $v_\phi$ shifts the spectrum to larger present-day comoving wavenumbers while leaving its peak height and shape nearly unchanged. 
This is because the Hubble-scale zero-temperature mass makes the $v_\phi$-dependence in the quadratic regime appear only through the expansion rate, which is canceled through Eqs.~\eqref{eq:zeta}, \eqref{eq:zetapowerspec}, and \eqref{eq:dimlesszetapowerspec}.

Figure~\ref{fig:powerspectrum} also shows that increasing $\xi$ suppresses the peak and moves the UV cutoff closer to it. A larger $\xi$ increases the magnitude of the tachyonic mass and accelerates the growth of $\langle\phi^2\rangle$. This increases the denominator in Eq.~\eqref{eq:zeta} and shortens the time required to reach $\phi_\times$.

Nevertheless, a moderate positive $\xi$ can enhance the curvature power spectrum compared to the minimally coupled case, for which Ref.~\cite{Lewicki:2021xku} finds $\mathcal P_\zeta\lesssim\mathcal O(10^{-3})$. The nonminimal term triggers the instability earlier, reaching $m_{\rm eff}^2 = 0$ by the $\xi$ contribution, which otherwise should rely on the cactus contribution. 
In the small $\xi$ limit, perturbatively introducing $\xi$ to the field evolution in the minimally coupled case shows that for $\xi$ satisfying
\beq
\xi \gg \frac{6g}{\pi} \l[ \frac{g^2}{16 \pi} \ln \l( \frac{8\pi \sqrt{2} M_{\rm P}}{g v_\phi} \r) \r]^2 \, , 
\eeq
the $\xi$ contribution dominates over the cactus contribution in $m_{\rm eff}^2$.
So for $\xi$ of order unity, the thermal fluctuations are relatively unsuppressed during the phase transition, changing the spectrum and its growth rate.
As a result, we have observationally allowed appreciable PBH production which we discuss in Sec.~\ref{sec:PBH}, and also larger scalar-induced GWs as in Sec.~\ref{sec:GW}.

Figure~\ref{fig:powerspectrumvaryingg} illustrates the dependence on $g$. Increasing $g$ raises $T_c$ through Eqs.~\eqref{eq:Tc0} and \eqref{eq:Tc}, making the thermal fluctuation at $t_c$ larger, enhancing the perturbation. But this effect is largely canceled in Eq.~\eqref{eq:zetapowerspec} by the increase of the growth rate, so the peak height nearly remains constant while the peak position moves toward larger $k$, bringing the UV cutoff closer to the peak. Only for $g \gtrsim 0.5$, does the cactus contribution affect the peak height. Combined with the $v_\phi$ dependence above, the peak height is almost independent of both parameters, $g$ and $v_\phi$, for $g \lesssim \mathcal{O}(0.1)$, while only the peak position changes according to $H_{\rm c}$.

\section{Phenomenology} \label{sec:pheno}

As we found $\mathcal{P}_{\zeta, k}$ can reach $\mathcal{O} (10^{-1} \text{--} 10^{-2})$, we now analyze the associated PBH production and scalar-induced GW signals. Cosmic strings formed by local $U(1)$ breaking provide an additional GW contribution, which can dominate at sufficiently large $v_\phi$.

\subsection{PBH formation} \label{sec:PBH}

We estimate PBH formation using the Press–Schechter formalism~\cite{Press:1973iz} and the Gaussian approximation in Sec.~\ref{ssec:curvpert}. For a Gaussian smoothing window, the comoving-gauge density contrast $\delta_R$ has variance~\cite{Young:2014ana}
\begin{equation}
    \sigma_R^2 = \left( \frac{4}{9} \right)^2 \int_0^\infty \frac{dk}{k} e^{-k^2 R^2} (kR)^4 \mathcal{P}_{\zeta, k} \, , \label{eq:sigmaRsq}
\end{equation}
where $R$ is the smoothing comoving scale. 
Then, the probability of having gravitational collapse at the scale $R$ is given by
\begin{equation}
    \beta_R = 2 \int^\infty_{\delta_c} \frac{e^{-\delta_R^2/2\sigma_R^2}}{(2\pi \sigma_R^2)^{1/2}} \, d\delta_R = \text{erfc} \left( \frac{\delta_c}{\sqrt{2}\sigma_R} \right) \, , \label{eq:beta}
\end{equation}
where $\delta_c \simeq 0.45$ is the collapse threshold~\cite{Musco:2020jjb}.

The PBH mass follows the mass-horizon relation
\begin{equation} \label{eq:MPBH}
    M_\text{PBH} = \gamma \frac{4\pi M_{\rm P}^2}{H_R} \approx \frac{32 \pi^2 \sqrt{2} \gamma M_{\rm P}^3}{g^2 v_\phi^2} e^{2 \Delta N_R} \, , 
\end{equation}
where $\gamma \simeq 0.2$ represents the fraction of horizon mass that collapses into a PBH~\cite{Carr:1975qj} and $H_R$ is the Hubble parameter when the collapse happens, i.e. when the associated scale reenters the horizon. 
For the approximation, we used
\beq\label{eq:HR}
    H_R \simeq H_{\rm c} e^{-2 \Delta N_R} 
\eeq
with $H_{\rm c}$ as the approximate expansion rate during the phase transition, and $\Delta N_R$ being the $e$-folds from the horizon exit of that scale to the end of the phase transition, assuming instantaneous reheating.
For an analytic estimate of the characteristic PBH mass, the dominating mode exits the horizon approximately at $t_c$, so taking $\Delta N_R\simeq\Delta N_{\rm fr}$ in \eq{eq:dN-fr} gives
\begin{equation} \label{eq:MPBH-Tc}
M_{\rm PBH} 
\approx \frac{32 \pi^2 \sqrt{2} \gamma M_{\rm P}^3}{g^2 v_\phi^2} \,  e^{\gamma_E/\bar{\nu}} \l( \frac{4 \pi}{g \ln^{1/2} \l( M_{\rm P}/v_\phi \r)} \r)^{2/\bar{\nu}} \, , 
\end{equation}
where we used $\phi_\times \sim v_\phi^2/M_{\rm P}$ in the argument of the log-factor.

Assuming instantaneous reheating followed by the standard thermal history, the formation probability is converted to the present PBH abundance $f_{\rm PBH} \equiv \Omega_{\rm PBH}/\Omega_{\rm c}$ by~\cite{Carr:2020gox}
\begin{equation} \label{eq:fPBH}
    f_{\rm PBH} \sim 4 \times 10^8 \gamma^{1/2} \beta_R \left( \frac{M_{\rm PBH}}{M_\odot} \right)^{-1/2} \, . 
\end{equation}
We evaluate $f_{\rm PBH}$ at each smoothing scale, and characterize the PBH population by the dominant scale in Table~\ref{tab:bench}. The exponential dependence of $\beta_R$ on $\sigma_R$ in Eq.~\eqref{eq:beta} sharpens the PBH spectrum relative to the curvature power spectrum, making it delta function-like with respect to the PBH mass. For PBHs that evaporate before the present epoch, $f_{\rm PBH}$ is the extrapolated abundance in the absence of evaporation.

\begin{figure}[ht]
    \centering
    \includegraphics[width=0.98\linewidth]{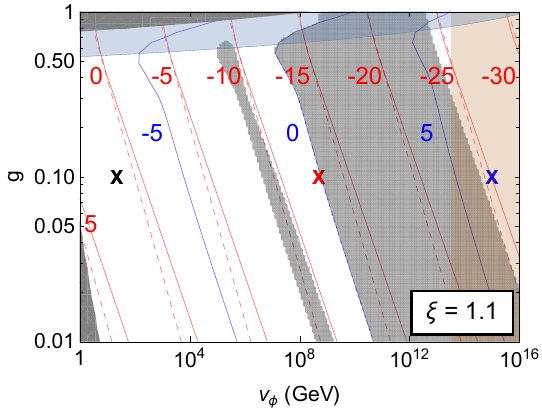}
    \includegraphics[width=0.98\linewidth]{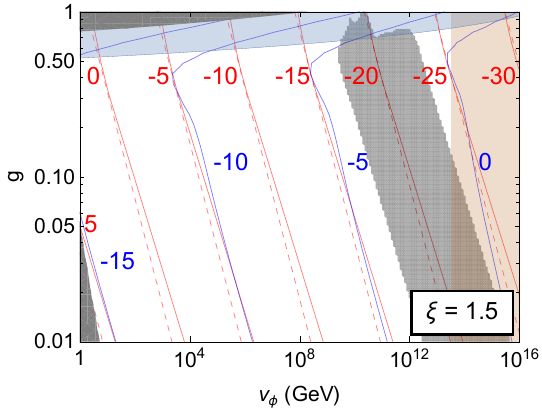}
    \caption{Contour plots of $M_{\rm PBH}$ (red; $\log_{10}(M_{\rm PBH}/M_{\odot})$; analytic estimation from Eq.~\eqref{eq:MPBH-Tc} is in red dashed) and $f_{\rm PBH}$ (blue; $\log_{10}(f_{\rm PBH})$) on $v$-$g$ plane, for $\xi = 1.1$ and $1.5$. The light-gray shaded regions are excluded by observational constraints.
    In both panels, the lower-left dark-gray shaded region has a reheating temperature of thermal inflation below $5 \, \text{MeV}$ and is excluded by the BBN requirement. The upper-left dark-gray shaded region lies outside the validity of the cactus approximation in Eq.~\eqref{eq:Tc}.
    The blue shaded region is the parameter space of first order phase transition.
    The orange shaded region is where the GW signal from the associated cosmic-strings could be constrained; see Sec.~\ref{sec:GW}.
    The three X markers in the upper panel indicate the GW benchmark points used in Fig.~\ref{fig:GW} and are listed in Table~\ref{tab:bench}. 
    }
    \label{fig:PBH}
\end{figure}

Figure~\ref{fig:PBH} shows $M_{\rm PBH}$ and $f_{\rm PBH}$ at the peak of PBH mass spectrum $f_{\rm PBH}(M_{\rm PBH})$, in the $(v_\phi,g)$ plane for $\xi=1.1$ and $1.5$. 
The nearly straight $M_{\rm PBH}$ contours reflect the leading scaling behavior $M_{\rm PBH}\propto 1/g^{2+2/\bar\nu}v_\phi^2$ in Eq.~\eqref{eq:MPBH-Tc}, decreasing toward larger $g$ and $v_\phi$.
The peak of $f_{\rm PBH} (M_{\rm PBH})$ is at the same scale as the peak of $\mathcal{P}_\zeta$, only with a slightly longer wavelength~\cite{Wang:2019kaf}, bringing Eq.~\eqref{eq:MPBH-Tc} into good agreement with the numerical result.

The behavior of $f_{\rm PBH}$ contours is more complicated. At small $g \lesssim \mathcal{O}(0.1)$, they are nearly parallel to $M_{\rm PBH}$ contours because the curvature peak and the formation probability depend little on $g$ and $v_\phi$, having $f_{\rm PBH}\propto M_{\rm PBH}^{-1/2}$ in Eq.~\eqref{eq:fPBH}. For larger $g \gtrsim 0.1$, the increasing height of the curvature peak increases $f_{\rm PBH}$, but even greater $g \gtrsim 0.5$ puts UV cutoff too close to the curvature peak (see Fig.~\ref{fig:powerspectrumvaryingg}), which decreases $f_{\rm PBH}$. These behaviors produce the bends in $f_{\rm PBH}$ contours. 
Had the full curvature power spectrum beyond our UV cutoff been taken into account, $f_{\rm PBH}$ in $g \gtrsim 0.5$ would have been less suppressed, resulting in smaller bending of its contours around $g \sim 0.5$.

All these parametric dependences are combined to give a broad producible PBH mass range in our scenario, $10^{-30} M_\odot \lesssim M_{\rm PBH} \lesssim 10^{5} M_\odot$. The mass function $f_{\rm PBH}$ can be above or below unity, and is highly sensitive to $\xi$, due to the exponential sensitivity of $\beta_R$ to the curvature power spectrum through Eqs.~\eqref{eq:sigmaRsq} and \eqref{eq:beta}.

The observational status of PBHs strongly depends on mass (see Ref.~\cite{Carr:2020gox} for a review). The nonevaporating population is mainly tested through gravitational effects, including microlensing, GWs, accretion signatures, and dynamical effects~\cite{Macho:2000nvd, Wilkinson:2001vv, EROS-2:2006ryy, Griest:2013esa, Griest:2013aaa, Niikura:2017zjd, Oguri:2017ock, Zumalacarregui:2017qqd, Niikura:2019kqi, Smyth:2019whb, Saito:2008jc, Ali-Haimoud:2017rtz, LIGOScientific:2019kan, Vaskonen:2019jpv, Chen:2019xse, Inoue:2017csr, Serpico:2020ehh, Lu:2020bmd, Monroy-Rodriguez:2014ula, Koushiappas:2017chw, Gong:2017sie, Carr:2018rid, MUSE:2020qbo}. 
These observations leave the DM mass window at $10^{-16} M_\odot \lesssim M_{\rm PBH} \lesssim 10^{-10} M_\odot$, bounded only by the overclosure.
Lighter PBHs are constrained by Hawking radiation, through CMB, $\gamma$- and cosmic-rays, radio signals, gas heating, and light-element abundances~\cite{Carr:2009jm, Acharya:2020jbv, Chluba:2020oip, Carr:2009jm, Carr:2016hva, Boudaud:2018hqb, Dasgupta:2019cae, DeRocco:2019fjq, Laha:2019ssq, Laha:2020ivk, Coogan:2020tuf, Chan:2020zry, Kim:2020ngi, Carr:2009jm}. 
Even lighter PBHs with $M_{\rm PBH}\lesssim10^{-24}M_\odot$ completely evaporated before Big Bang nucleosynthesis, and are difficult to constrain.

These constraints exclude the light gray shaded regions in Fig.~\ref{fig:PBH}. 
Since $f_{\rm PBH} \propto M_{\rm PBH}^{-1/2}$ in most of the parameter space (i.e., $g \lesssim 0.5$), the allowed and constrained bands approximately follow $M_{\rm PBH}$ contours. 
Viable PBH populations include PBH DM and SMBH seeds, where the latter can be in the mass range of $M_{\rm PBH} \gtrsim 50 M_\odot$ with $10^{-12} M_{\rm PBH}/M_\odot \lesssim f_{\rm PBH} \lesssim 10^{-8} M_{\rm PBH} / M_\odot$~\cite{Dimopoulos:2019wew}. Evaporating PBHs may be excluded by the associated cosmic-string GWs (orange shaded region; see the next subsection), but it depends on the string model and cosmological history. 
The constrained bands sensitively depend on $\xi$, and widen with decreasing $\xi$. For example, the DM window is closed for $\xi \lesssim 1$, and another constrained band appears in the mass range of SMBH seeds.

\begin{figure}[ht]
    \centering
    \includegraphics[width=0.98\linewidth]{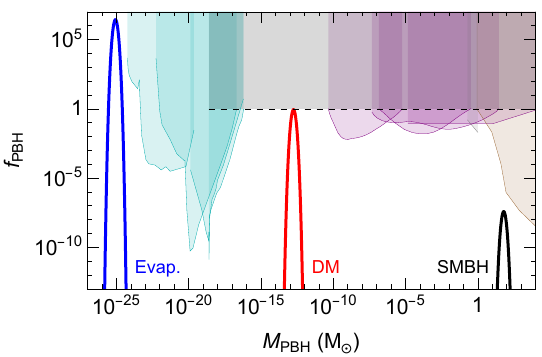}
    \caption{
    PBH abundance spectra for the three benchmark points, ``SMBH", ``DM", and ``Evap." in Table~\ref{tab:bench} and the upper panel of Fig.~\ref{fig:PBH}. Overlaid constraints are from BBN, CMB spectral distortions and anisotropies, extragalactic and galactic backgrounds (cyan)~\cite{Carr:2009jm, Carr:2016hva, Boudaud:2018hqb, Acharya:2020jbv, Chluba:2020oip}, SUBARU, EROS, OGLE, Icarus (magenta)~\cite{EROS-2:2006ryy, Oguri:2017ock, Niikura:2017zjd, Niikura:2019kqi}, binary merger rates (gray)~\cite{LIGOScientific:2019kan}, and accretion-induced CMB anisotropies (brown)~\cite{Serpico:2020ehh}.}
    \label{fig:fPBHConstraint}
\end{figure}

We pick three benchmark points listed in Table~\ref{tab:bench} and present them with three X markers in the upper panel of Fig.~\ref{fig:PBH}. The black, red, and blue markers correspond to the parameter points that produce SMBH seeds (``SMBH''), PBH DM (``DM''), and evaporating PBHs (``Evap.''), respectively. 
Their PBH abundance spectra are plotted in Fig.~\ref{fig:fPBHConstraint}, along with observational constraints.

\subsection{Gravitational waves} \label{sec:GW}

Curvature perturbations that are sufficient to produce PBHs also induce significant secondary scalar-induced GWs. 
Their present-day spectrum is given by~\cite{Espinosa:2018eve, Kohri:2018awv, Inomata:2019yww}
\begin{align}
    &\Omega_{\rm GW}^{\rm SI} (f) h^2 \nonumber \\
    &\quad\simeq 4\times 10^{-4} \times \left(\frac{100}{g_{*s}} \right)^{1/3} \left(\frac{g_*}{g_{*s}} \right) \nonumber \\
    & \quad\quad\times \int^1_{-1} dx \int^\infty_1 dy \, \mathcal{P}_{\zeta, (x+y) \pi f} \mathcal{P}_{\zeta, (y-x)\pi f} \mathcal{J}^2(x,y) \, , 
\end{align}
where
\begin{align}
    \mathcal{J}^2(x, y) ={}& \frac{Z^2 (X+2)^2 (Y+2)^2}{(X - Y)^8} \nonumber \\
    & \times \left[ \left (X - Y + \frac{Z}{2} \ln \left| \frac{Y}{X} \right| \right)^2 + \frac{\pi^2 Z^2}{4} \Theta(Y) \right] 
\end{align}
with
\begin{equation}
    X = x^2 - 3 \, , \quad Y = y^2 - 3 \, , \quad Z = X+Y \, , 
\end{equation}
and $\Theta$ is the Heaviside step function. 

Another source of GWs in our scenario is cosmic strings from local $U(1)$ breaking.
Once the cosmic string network enters the scaling regime, oscillating loops emit GWs with the spectrum~\cite{Gouttenoire:2019kij}
\begin{equation}
    \Omega_{\rm GW}^{\rm CS} (f) =  \sum_j \Omega_{\rm GW}^j (f) \, ,
\end{equation}
where 
\begin{align}
    \Omega_{\rm GW}^j (f) ={}& \frac{1}{\rho_c} \frac{2 j}{f} \frac{\mathcal{F}_\alpha \Gamma^{(j)} G \mu^2}{\alpha (\alpha + \Gamma G \mu)} \nonumber \\
    &\times \int^{t_{\rm now}}_{t_{\rm osc}} d \tilde{t} \, \frac{C_{\rm eff}(t_i)}{t_i^4} \frac{a^2(\tilde{t}) a^3(t_i)}{a^5(t_{\rm now})} \Theta(t_i - t_{\rm osc})
\end{align}
is the contribution of $j$-th oscillation mode of loops, $\rho_c$ is the present-day critical density, $\alpha\simeq 0.1$ is the initial ratio of the loop length to the horizon size, $\mathcal{F}_\alpha \simeq 0.1$ is the fraction of loops formed with the length ratio $\alpha$, $\Gamma \simeq 50$ is the gravitational loop-emission efficiency, $\Gamma^{(j)} \simeq j^{-4/3} \Gamma / 3.60$ is the efficiency contribution of $j$-th mode in the case of cusps as an example, $\mu \simeq \pi v_\phi^2$ is the tension of the cosmic string, $C_{\rm eff} (t) \simeq 5.4 \, \text{(RD) or } 0.39 \, \text{(MD)}$ is the loop-production efficiency, $t_{\rm osc}$ is the time at which the oscillation starts ($t_{\rm osc} \approx t_{\rm end}$ for our case with neglecting the effect of friction), and
\begin{equation}
    t_i = t_i (f, \tilde{t}) = \frac{1}{\alpha + \Gamma G \mu} \left( \frac{2j}{f} \frac{a(\tilde{t})}{a(t_{\rm now})} + \Gamma G \mu \tilde{t} \right)
\end{equation}
is the loop formation time with ${\tilde t}$ being the time of GW emission. 
Also included was the suppression of high-frequency GWs by particle emission~\cite{Gouttenoire:2019kij}.

\begin{figure}[ht]
    \centering
    \includegraphics[width=0.98\linewidth]{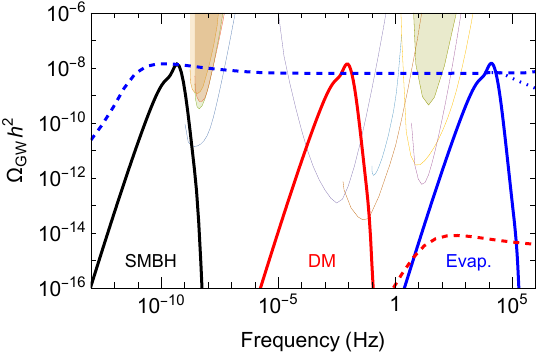}
    \caption{GW spectra for the three benchmark points in Table~\ref{tab:bench}. Solid curves show the scalar-induced GWs and dashed curves are for cosmic-string contribution, with cusp-induced particle emission is taken into account in dotted curve.
    Overlaid are the power-law integrated sensitivity curves~\cite{Thrane:2013oya, Schmitz:2020syl, Caprini:2024ofd} of SKA~\cite{Janssen:2014dka}, EPTA~\cite{EPTA:2015qep}, PPTA~\cite{Shannon:2015ect}, NANOGrav~\cite{NANOGRAV:2018hou}, LISA~\cite{LISA:2017pwj, Babak:2021mhe}, AEDGE~\cite{AEDGE:2019nxb}, AION-km~\cite{Badurina:2019hst}, ET~\cite{Sathyaprakash:2012jk, Danilishin:2023ET}, CE~\cite{Reitze:2019iox, Evans:2021gyd, CosmicExplorer:T2000017}, and LIGO~\cite{KAGRA:2013rdx}. Observational constraints are distinguished by shading.}
    \label{fig:GW}
\end{figure}

Fig.~\ref{fig:GW} shows GW spectra of scalar-induced GWs (solid) and cosmic-string GWs (dashed) for the three benchmark points listed in Table~\ref{tab:bench}. 
A useful estimate relates the characteristic frequency of the scalar-induced GW to the PBH mass as~\cite{Kohri:2020qqd} 
\begin{equation}
    f_{\rm peak} \simeq 5\times 10^{-9} \, \text{Hz} \times \left(\frac{M_{\rm PBH, \rm peak}}{M_\odot} \right)^{-1/2} \, ,
\end{equation}
which can be checked from Table~\ref{tab:bench} and Fig.~\ref{fig:GW}.
The peak heights are almost the same for all three benchmarks, following their curvature spectra amplitudes (Fig.~\ref{fig:powerspectrum}).

The ``SMBH'' and ``DM'' benchmarks provide interesting observational targets. 
The ``SMBH'' signal peaks at frequencies of order $10^{-9}\,\mathrm{Hz}$, overlapping the projected SKA sensitivity range. 
Increasing $v_\phi$ shifts the GW peak toward the frequency ranges of EPTA, PPTA, and NANOGrav, but reduces the associated PBH mass below the mass range for SMBH seeds. 
Hence, the SMBH seeds in our scenario cannot be responsible for the GW signal discovered at NANOGrav~\cite{NANOGrav:2023gor}.
Meanwhile, the ``DM'' signal peaks near $10^{-2}\,\mathrm{Hz}$, deep in the projected sensitivity ranges of LISA and AEDGE, possibly being a target of those experiments.

The cosmic-string GW has a broad plateau whose height depends on the string tension and hence on $v_\phi$. 
The signal could reach the PTA and LIGO sensitivity ranges for high $v_\phi$ in Fig.~\ref{fig:GW}, which gives the orange shaded region in Fig.~\ref{fig:PBH}.
The cosmic-string GW spectrum depends on the UV embedding of the $U(1)$ symmetry and on the cosmological history. 
For example, cosmic strings could be metastable by quantum nucleation of monopole and antimonopole pairs~\cite{Langacker:1980kd}, and an early matter-dominated era can alter the GW spectrum at high frequencies. 
These effects may weaken the string constraints and leave the orange shaded region in Fig.~\ref{fig:PBH} unconstrained, but a modified cosmological history must be incorporated consistently into the PBH and GW predictions.

\section{Summary and discussion} \label{sec:summary}

In this work, we considered a scale-invariant local $U(1)$ model and showed that scale-invariance can naturally produce large cosmological perturbations for seeding PBHs. 

Theories based on scale invariance are likely to have quadratic nonminimal scalar–curvature couplings to generate the Planck scale.
For subPlanckian scalars, such an interaction induces a mass term in the Einstein-frame potential. In particular, positive nonminimal coupling $\xi \sim \mathcal{O}(1)$ generates a negative Hubble-scale mass term. 
This changes the nature of the exit from thermal inflation driven by a CW potential of the Higgs field, relative to the minimally coupled case~\cite{Lewicki:2021xku}.

Thermal inflation with a negative Hubble-scale mass term was previously considered in the context of supergravity~\cite{Dimopoulos:2019wew}. 
Our scenario provides a nonsupersymmetric realization of this mechanism, based on the scale invariance, and has several major differences in comparison to earlier related works~\cite{Dimopoulos:2019wew,Lewicki:2021xku}. 
First, the total duration of thermal inflation only mildly depends on the gauge coupling $g$, and can remain small enough for $g \ll 1$, for a wide range of the symmetry-breaking scale $v_\phi$. 
Second, the induced mass term triggers a second order phase transition when thermal fluctuations dominate over vacuum fluctuations, producing sufficiently large density perturbations to have PBH formation. 
Third, the very weak dependence of the curvature power spectrum on $g$ and $v_\phi$ allows a broad range of producible PBH mass, including SMBH seeds and PBH DM, with associated scalar-induced GW signals at different frequency bands.
The model also permits light evaporating PBHs, under possible constraint by cosmic-string GWs, depending on the string model and cosmological history.

Although the association between large curvature perturbations, PBHs, and scalar-induced GWs is generic, the detailed GW spectrum can contain information about the fast-roll potential. 
In particular, the spectrum above the peak frequency may distinguish the present scenario from related models~\cite{Dimopoulos:2019wew,Lewicki:2021xku}.
We leave a detailed study of these features to future work.

\begin{acknowledgments}
W.I.P. thanks Tomasz Dutka for his contribution during the early stages of this work. 
T.H.K. is supported by the KIAS Individual Grant PG095202 at the Korea Institute for Advanced Study.
W.I.P. was supported by Global - Learning \& Academic research institution for Master’s·PhD students, and Postdocs (LAMP) Program of the National Research Foundation of Korea (NRF) grant funded by the Ministry of Education (RS-2024-00443714).
\end{acknowledgments}

\appendix

\section{Analytic estimate of the Euclidean action}
\label{app:PT}

We consider the finite-temperature effective potential introduced in \eq{eq:VT-eff}, written as
\begin{align}
V(\phi, T) &\approx V_0 - \frac{\pi^2}{10} T^4 + \frac{1}{2} M^2 \phi^2
\nonumber \\
& - \frac{1}{3} ET \phi^3 + \frac{1}{4} \lambda_{\rm CW} \phi^4 \l[ \ln \l( \frac{\phi^2}{v_\phi^2} \r) - \frac{1}{2} \r] \, ,
\end{align}
where
\begin{align}
V_0 &= \frac{1}{8} \lambda_{\rm CW} v_\phi^4
\\
M^2 &= \frac{g^2}{4} \l( 1 - \frac{\bar{\lambda}_{\rm c}}{g^2} \r) \l( T^2 - T_{\rm c}^2 \r) 
\nonumber \\
&= \frac{g^2}{4} \l( 1 - \frac{\bar{\lambda}_{\rm c}}{g^2} \r) \ep \l( \ep + 2 \r) T_{\rm c}^2
\\
E &= \frac{3g^3}{4 \pi}
\\
\lambda_{\rm CW} &= \frac{3 g^4}{16 \pi^2}
\end{align}
with $T = \l( \ep + 1 \r) T_{\rm c}$, and 
\begin{align}
T_{\rm c} &\equiv \sqrt{\frac{1+\bar{\lambda}_{\rm c}/8\pi g \xi}{1-\bar{\lambda}_{\rm c}/g^2}} \times \frac{\sqrt{24 \xi} H_{\rm c}}{g} \, ,
\\
H_{\rm c} &\approx \frac{g^2 v_\phi^2}{8 \pi \sqrt{2} M_{\rm P}} \, .
\end{align}
For a symmetry-breaking phase transition, the flat spacetime 3D Euclidean action $S_3$ at finite temperature is given by
\beq \label{eq:S3}
S_3 = 4 \pi \int dr \, r^2 \l[ \frac{1}{2} \l( \frac{d\phi}{dr} \r)^2 + V_{\rm eff} \r] \, , 
\eeq
where
\beq
V_{\rm eff} \equiv V(\phi, T) - V(0, T) \, .
\eeq
For a first order phase transition, the bounce equation is
\beq \label{eq:bounce}
0 = \frac{d^2 \phi}{d r^2} + \frac{2}{r} \frac{d\phi}{dr} - \frac{d V_{\rm eff}}{d\phi}
\eeq
with the boundary conditions 
\begin{align}
\l. \frac{d\phi}{dr} \r|_{r=0} &= 0 
\\
\phi(r\to \infty) &= 0 \, ,
\end{align}
where $r$ is the distance from the center of a bubble at nucleation.
To obtain an analytic estimate of the tunneling action, we use a sech-function ansatz for the bubble profile, 
\beq \label{eq:sech-bounce}
\phi_{\rm b}(r) = \frac{\phi_{\rm e}}{\cosh{(r/R)}} \, , 
\eeq
where $\phi_{\rm e}$ and $R$ stand for the exit-point and the characteristic radius of the bubbles at nucleation.
Substituting Eq.~\eqref{eq:sech-bounce} into Eq.~\eqref{eq:S3} gives
\beq \label{eq:S3-from-sech}
\frac{S_3}{4 \pi R \phi_{\rm e}^2} 
\simeq K + P_2 R^2 M^2 - P_3 ETR^2 \phi_{\rm e} - P_4 \bar{\lambda} R^2 \phi_{\rm e}^2 \, ,
\eeq
where
\begin{align}
\bar{\lambda} &\equiv \lambda_{\rm CW} \l[ \ln \l( \frac{v_\phi^2}{\phi_{\rm e}^2} \r)+\frac{1}{2} - \frac{P_L}{P_4}\r]
\\
K &= \frac{\pi^2 + 12}{72}
\\
P_2 &= \frac{\pi^2}{24}
\\
P_3 &= \frac{\pi \l( \pi^2-8 \r)}{48}
\\
P_4 &= \frac{\pi^2-6}{72}
\\
P_L &= \frac{\pi^2 \l( 6 \ln2 - 5 \r) + 6 \l( 11-6\ln2-6\zeta(3) \r)}{216} \, .
\end{align}
Then, applying the variational method, one finds 
\begin{align}
R &= \frac{1}{M} \times q(\beta)
\\
\phi_{\rm e} &= \frac{ET}{\bar{\lambda}} \times h(\beta) 
\end{align}
where
\begin{align}
\beta &\equiv \frac{\l( P_3 ET \r)^2}{P_2 P_4 \bar{\lambda} M^2} 
\nonumber \\
&= \frac{6P_3^2}{P_2 P_4} \, \frac{(\epsilon+1)^2}{\epsilon (\epsilon+2)} 
\nonumber \\
&\quad \times \frac{1}{\l( 1-\frac{\bar{\lambda}_{\rm c}}{g^2} \r) \l[ \ln\l( \frac{2\pi M_{\rm P}}{\sqrt{3 \xi}v_\phi}  \r) + \frac{1}{4} - \frac{P_L}{2P_4} \r]}
\\
q(\beta) &= \l[ \l( \frac{K}{2P_2} \r) \frac{1+\l( 8/3\beta \r) + \sqrt{1+32/9\beta}}{1+4/\beta} \r]^{1/2}
\\
h(\beta) &= \frac{3P_3}{4P_4} \times \l[ \sqrt{1+32/9\beta}-1 \r] \, .
\end{align}
Here, we replaced $\phi_e$ with $2T_{\rm c0}/g$ in the log function, which significantly simplifies the calculation without altering the final estimate for $S_3 / T$.
The resulting action is
\begin{align}
\frac{S_3}{T}
&\simeq \frac{8 \pi}{3} \l( \frac{E}{\bar{\lambda}} \r)^2 \l( \frac{T}{M} \r) \times K q h^2
\nonumber \\
&\simeq \frac{64 \pi^3 \l( \ep + 1 \r) \times K q h^2}{3 g^3 \l( 1- \frac{\bar{\lambda}_{\rm c}}{g^2} \r)^{\frac{1}{2}} \l( \ep \l( \ep + 2 \r) \r)^{\frac{1}{2}} \l[ \ln \l( \frac{g v_\phi}{2T_{\rm c0}} \r) + \frac{1}{4} \r]^2} \, .
\end{align}
For $\bar{\lambda}_{\rm c} \ll g^2$ which will be of our interest, $\beta < 1$ gives $q(\beta) \sim 1/2$ and $h(\beta) \propto \beta^{-1/2}$.
Also, for $\ep \ll 1$, one finds $\beta \propto 1/\ep$.
Thus, the qualitative scaling is
\beq
\frac{S_3}{T} \propto \frac{\l[ \ep \l( \ep+2 \r) \r]^{1/2}}{g^3 \l(\ep+1 \r)}
\xrightarrow[]{\beta , \ep \ll 1} \frac{\ep^{1/2}}{g^3} \, .
\eeq

The Kibble-Zurek mechanism~\cite{Kibble:1976sj,Zurek:1985qw,Zurek:1996sj} indicates that the crossover between first-order and second-order phase transitions takes place when the relaxation time $\tau_R$ becomes equal to the quenching time $\tau_Q$. 
We find
\begin{align}
\tau_{\rm R} &= \frac{1}{M} = \frac{1}{\l[ 6 \xi \l( 1+ \frac{\bar{\lambda}_{\rm c}}{8 \pi g \xi} \r)\r]^{1/2} \sqrt{\ep \l( \ep+2 \r)} H} \, ,
\\
\tau_{\rm Q} &\equiv \frac{\ep}{\l| \dot{\ep} \r|} = \frac{\ep}{(\ep+1)H} \, .
\end{align}
Hence, $\tau_{\rm R}(\ep_*) = \tau_{\rm Q}(\ep_*)$ gives 
\beq \label{eq:quenching}
\frac{\ep_* \l( \ep_*+2 \r)^{1/3}}{\l( \ep_*+1 \r)^{2/3}} = \l[ 6 \xi \l( 1+ \frac{\bar{\lambda}_{\rm c}}{8 \pi g \xi} \r)\r]^{-1/3} \, .
\eeq
For $\ep>\ep_*$, a first-order phase transition can take place if 
\beq
\frac{S_3}{T} \approx \l( \frac{S_3}{T} \r)_{\rm c} \equiv 4 \ln \l( \frac{T}{H} \r) + \frac{3}{2} \ln \l( \frac{S_3}{2\pi T} \r)_{\rm c} \, .
\eeq 
On the other hand, $\epsilon < \epsilon_*$ predicts a second-order phase transition even if the condition for a first-order phase transition is satisfied.
Since $\ep(T)$ decreases from a large value, the criterion to avoid a first-order transition for $\ep>\ep_*$ is $S_3/T > \l( S_3/T \r)_{\rm c}$. Figure~\ref{fig:PT-para-space} shows this separation of phases.
Compared to the minimally coupled case, the induced mass term shifts the estimated boundary toward larger $g$ at fixed $v_\phi$; this term reduces $\ep_*$ in comparison to the minimally coupled case where a decrease of $\ep_*$ implies an increase of $g$.

\section{Evolution of power spectrum of $\phi$} 
\label{app:phipowerspectrum}

We derive a generalization of Eq.~\eqref{eq:PkEOM} that includes thermal noise and dissipation, with time-dependent coefficients. The derivation assumes mode separation and Gaussian white noise. Its microscopic thermal interpretation additionally requires a sufficiently shorter equilibration time scale in comparison to the dynamic time scale of the field. Subject to these assumptions, the Langevin equation is~\cite{Berera:2008ar, Graham:2009bf, Hiramatsu:2014uta, Bastero-Gil:2023sub, Dutka:2025oqt}
\begin{equation}
    \ddot{\phi}_{\vec{k}} (t) + b(t) \dot{\phi}_{\vec{k}} (t) + q^2(t) \phi_{\vec{k}} (t) = \xi_{\vec{k}} (t) \, , \label{eq:Langevin}
\end{equation}
where $b(t)$ is a damping coefficient, $q^2(t)$ is the effective frequency squared, and $\xi_{\vec{k}}$ is a stochastic noise with correlation function
\begin{equation}
    \langle \xi_{\vec{k}} (t) \, \xi_{\vec{k}'} (t')\rangle = D(t) \delta^3(\vec{k} + \vec{k}') \, \delta(t-t') \, .
\end{equation}

Eq.~\eqref{eq:Langevin} can be decomposed into a system of two first-order differential equations. In differential form, they are
\begin{subequations}
\begin{align}
    d\phi_{\vec{k}} &= \pi_{\vec{k}} dt \, ,  \\
    d\pi_{\vec{k}} &= (-b \pi_{\vec{k}} - q^2 \phi_{\vec{k}})dt + \sqrt{D} dW_{\vec{k}} \, , 
\end{align}
\end{subequations}
where $dW_{\vec{k}}$ is the infinitesimal Wiener increment which satisfies 
\begin{equation}
    \langle dW_{\vec{k}} \rangle = 0 \, , \quad \langle dW_{\vec{k}} \, dW_{\vec{k}'} \rangle = \delta^3 (\vec{k} + \vec{k}') dt \, . \label{eq:dW}
\end{equation}
We then obtain a system of first-order differential equations for $\langle \phi_{\vec{k}} \phi_{-\vec{k}} \rangle$, $\langle \phi_{\vec{k}} \pi_{-\vec{k}} \rangle$, and $\langle \pi_{\vec{k}} \pi_{-\vec{k}}\rangle$. As $\phi$ is a real field, we have $\phi_{-\vec k} = \phi_{\vec k}^*$ and $\pi_{-\vec k} = \pi_{\vec k}^*$, while statistical isotropy implies $\langle \phi_{\vec k} \pi_{-\vec k} \rangle = \langle \phi_{-\vec k} \pi_{\vec k} \rangle$, which makes all three correlators real. Keeping terms up to first order in $dt$ with $dW_{\vec{k}} \sim \mathcal{O} (dt^{1/2})$, taking the ensemble average gives~\cite{mendez2011instabilities}
\begin{subequations}
\begin{align}
    \frac{d}{dt} \langle \phi_{\vec{k}} \phi_{-\vec{k}} \rangle &= 2 \langle \phi_{\vec{k}} \pi_{-\vec{k}} \rangle \\
    \frac{d}{dt} \langle \phi_{\vec{k}} \pi_{-\vec{k}} \rangle &= \langle \pi_{\vec{k}} \pi_{-\vec{k}} \rangle - b \langle \phi_{\vec{k}} \pi_{-\vec{k}} \rangle - q^2 \langle \phi_{\vec{k}} \phi_{-\vec{k}} \rangle \\
    \frac{d}{dt} \langle \pi_{\vec{k}} \pi_{-\vec{k}} \rangle &= -2b \langle \pi_{\vec{k}} \pi_{-\vec{k}} \rangle - 2 q^2 \langle \phi_{\vec{k}} \pi_{-\vec{k}} \rangle + D \delta^3 (\vec{0}) 
\end{align}
\end{subequations}
where $\delta^3 (\vec{0})$ comes from $\vec{k}' = -\vec{k}$ in Eq.~\eqref{eq:dW}. 
Eliminating $\langle \phi_{\vec{k}} \pi_{-\vec{k}}\rangle$ and $\langle \pi_{\vec{k}} \pi_{-\vec{k}}\rangle$ results in
\begin{align}
    &\frac{d^3}{dt^3} \langle \phi_{\vec{k}} \phi_{-\vec{k}} \rangle + 3 b \frac{d^2}{dt^2} \langle \phi_{\vec{k}} \phi_{-\vec{k}} \rangle \nonumber \\
    & \quad + (\dot{b} + 2 b^2 + 4q^2) \frac{d}{dt} \langle \phi_{\vec{k}} \phi_{-\vec{k}} \rangle \nonumber \\
    & \quad + [2 \dot{(q^2)} + 4 b q^2] \langle \phi_{\vec{k}} \phi_{-\vec{k}} \rangle = 2 D \delta^3 (\vec{0}) \, .
\end{align}
For the power spectrum defined in Eq.~\eqref{eq:powerspectrumdefine}, $\langle \phi_{\vec{k}} \phi_{-\vec{k}} \rangle = P_k \delta^3 (\vec{0})$, and we obtain the equation for $P_k$ as
\begin{equation}
    \dddot{P}_k + 3 b \ddot{P}_k + (\dot{b} + 2 b^2 + 4q^2) \dot{P}_k + [2 \dot{(q^2)} + 4 b q^2] P_k = 2 D \, . \label{eq:PkEOMGeneral}
\end{equation}

We performed two checks of Eq.~\eqref{eq:PkEOMGeneral}. For time-independent coefficients and vanishing initial field and velocity fluctuations, it reproduces the analytic solution of Ref.~\cite{Dutka:2025oqt}. For our model, assuming the Langevin description applies, the numerical solution agrees with the thermal equilibrium solution $P_k=(2n_k+1)|\chi_k|^2/a^2$ at $T\gg T_c$ and evolves smoothly into the tachyonic regime across $T = T_c$. We expect Eq.~\eqref{eq:PkEOMGeneral} to have a broader usage in other linear stochastic systems satisfying the stated assumptions.

\bibliography{references}

\end{document}